\documentclass[12pt]{article}

\usepackage{a4wide}
\usepackage{amssymb}
\usepackage{bm}
\usepackage{latexsym}
\usepackage{dcolumn}
\usepackage{amsfonts,amssymb}
\usepackage{graphicx,epsfig}
\usepackage{psfrag}
\usepackage{amsmath,amssymb}
\usepackage{verbatim}

\usepackage{bm}
\usepackage{latexsym}
\usepackage{mathrsfs}
\usepackage[T1]{fontenc}

\newcommand{\be}{\begin{equation}}
\newcommand{\ee}{\end{equation}}
\newcommand{\ba}{\begin{eqnarray}}
\newcommand{\ea}{\end{eqnarray}}
\newcommand{\ban}{\begin{eqnarray*}}
\newcommand{\ean}{\end{eqnarray*}}

\newcommand{\detailtexcount}{%
  \immediate\write18{texcount -merge -sum -incbib -dir \jobname.tex > \jobname.wcdetail }%
  \verbatiminput{\jobname.wcdetail}%
}

\newcommand{\quickwordcount}{%
  \immediate\write18{texcount -1 -sum -merge \jobname.tex > \jobname-words.sum }%
  \input{\jobname-words.sum} words%
}

\newcommand{\quickcharcount}{%
  \immediate\write18{texcount -1 -sum -merge -char \jobname.tex > \jobname-chars.sum }%
  \input{\jobname-chars.sum} characters (not including spaces)%
}

\begin{document}
\begin{titlepage}
\pagestyle{empty}
\baselineskip=21pt
\vspace{2cm}
\begin{center}
{\bf {\Large 
Is there an upper bound on the size of a black-hole?\footnote{Essay received Honorable mention in Gravity Research Foundation essay competition-2018}
}}
\end{center}
\begin{center}
\vskip 0.2in
{\bf Swastik Bhattacharya}$^{(a)}$ and {\bf S. Shankaranarayanan}$^{(b)}$
\vskip 0.1in
$^{(a)}$ {\it Department of Physics, BITS Pilani Hyderabad, Hyderabad 500078, Telangana State, India.} 

$^{(b)}$ {\it Department of Physics, Indian Institute of Technology Bombay, Mumbai 400076, India} \\
{\tt Email: swastik@hyderabad.bits-pilani.ac.in , shanki@phy.iitb.ac.in}\\
\end{center}

\vspace*{0.5cm}

\begin{abstract}
According to the third law of Thermodynamics, it takes an infinite number of steps for any object, including black-holes, to reach zero temperature. For any physical system, the process of cooling to absolute zero corresponds to erasing information or generating pure states. In contrast with the ordinary matter, the black-hole temperature can be lowered only by adding matter-energy into it. However, it is impossible to remove the statistical fluctuations of the infalling matter-energy. The fluctuations lead to the fact the black-holes have a finite lower temperature and, hence, an upper bound on the horizon radius. We make an estimate of the upper bound for the horizon radius which is curiosly comparable to Hubble horizon. We compare this bound with known results and discuss its implications.
\end{abstract}

\vspace*{2.0cm}

\end{titlepage}

\baselineskip=18pt

Black-hole entropy has remained one of the most inexplicable quantities in Theoretical Physics~\cite{entropyrev}. It is puzzling because it raises two related issues:  Entropy is very large for macroscopic black-holes --- it is the highest for objects of similar size~\cite{Entrobound}, and it lacks any understanding from a statistical mechanical description~\cite{entropyrev}. Although the two issues may seem different, they are two sides of the same coin. Apart from the fact that the black-hole entropy comes from the lack of information about the microscopic degrees of freedom of space-time inside the event-horizon, little else is agreed upon in the literature~\cite{entropyrev}. 

There have been several proposals to interpret black-hole entropy~\cite{entropyrev}. One proposal states that the black-hole entropy is the number of ways a black-hole (characterized by Mass $M$, Charge $Q$ and angular momentum $a$) is formed~\cite{Thorne}. Thus, the increase in the black-hole entropy corresponds to the number of ways in which the matter-energy can fall into the black-hole. For a faraway observer, the black-hole entropy corresponds to not being to able determine how the microscopic degrees of freedom (DOF) fell into the black-hole. {\it A very important consequence of this proposal is that  the information about the black-hole DOF also resides in the information about how the matter-energy fell into the black-hole.} However, the distant observer can receive the Hawking radiation from the black-holes. But as this is thermal, it gives very little information about the black-hole DOF.

Thus two opposite processes take place in the presence of black-holes. First, black-hole entropy keeps increasing as it swallows matter around it leading to the reduction in black-hole temperature. Second, due to Hawking radiation black-hole reduces its mass and hence raising its temperature. It is often assumed, in the literature, that the second process dominates and one is lead to the question: What happens at the end-stages of the black-hole evaporation? In this essay, we ask what happens when the first process dominates: Do black-holes keep increasing {\sl ad infinitum}? Is there an upper bound on the size of a black-hole? We address these questions in this essay. We first argue, using general assumptions, that an upper bound exists. Later, within the Horizon-Fluid correspondence~\cite{FluidGrav,FluidGrav2,FluidGrav3,FluidGrav4}, we obtain a 
value of this bound. 

The third law of black-hole mechanics states that it takes infinite steps to reach a black-hole with zero temperature. More generally, for any physical system, the process of cooling to absolute zero corresponds to erasing information or generating pure states~\cite{Masanes-Oppenheim}. In the case of black-holes, the physical process of decreasing the temperature corresponds to adding energy to the system. However, the statistical fluctuations of the infalling matter-energy can never be removed. The fluctuations lead to the fact there exist a non-zero lower bound on the black-hole temperature and, hence, an upper bound on the horizon radius. 

The above discussion was general and qualitative. In the rest of this essay, we obtain the upper-bound using Fluid-Gravity correspondence following Damour~\cite{FluidGrav}. Fluid-Gravity correspondence aims to associate fluid degrees of freedom to the horizon and, eventually, to the gravitational degrees of freedom \cite{FluidGrav,FluidGrav2}. Fluid-Gravity correspondence can be considered as an extension of black-hole thermodynamics, where charges are upgraded into local currents and black-hole entropy into a local entropy current. Since Damour's calculation 30 years ago~\cite{FluidGrav}, attempts have been made to gain new physical insight from the Fluid-Gravity correspondence. More importantly, how to use the Fluid-Gravity correspondence to connect macroscopic and microscopic physics through the study of the statistical properties of the fluid on the black-hole horizon. Recently, the present authors have shown that horizon-fluid is of physical interest~\cite{FluidGrav3}, and obtained several physical quantities using general assumptions about the horizon-fluid~\cite{FluidGrav4}.

To obtain the bound, we define $N$ as the number of degrees of freedom (DOF) of the horizon-fluid that is proportional to the area of the horizon (or entropy). $N$ is the measure of the number of Planck scale DOF on the horizon~\cite{Wheeler}. We also define the number density ($N$ per black-hole area) as a field $\rho$. At the semi-classical level, $\rho$ is constant. However, the field can have fluctuations about the mean value. For a given temperature, one can determine the average energy ($\bar{\epsilon}$) of the horizon DOF~\cite{PaddyEquipart}. Let us compare this energy to the infalling matter-energy constituents that increase the energy and entropy of the black-hole. It is known that only the highest energy excitations contribute to the entropy~\cite{BrickWall}. We assume the highest energy scale to be the same for all the fields. Thus, the horizon can be viewed as surrounded by a thin layer of such excitations. Even for a macroscopic black-hole, $\bar{\epsilon}$ is much higher than the excitations of the surrounding medium. However as the temperature of the black-hole decreases, $\bar{\epsilon}$ decreases. Eventually, for a specific value of the black-hole temperature, say $T=T^c$, the above two energy scales become comparable. Unlike the cooling of ordinary objects, we cannot decrease the energy of the excitations of the medium surrounding the black-hole. Hence, the black-hole temperature cannot be lowered beyond $T^c$. For any lower value of $\bar{\epsilon}$, the excitations of the surrounding medium would impart energy increasing $\bar{\epsilon}$ and the black-hole temperature. 

Thermal wave-length $({\lambda_T})$ corresponding to the average energy ($\bar{\epsilon} = k_B T/2$) for the field $\rho$ is given by:
\begin{equation}
\lambda_T \sim \frac{h \, v}{k_B \, T}. \label{TLambda}
\end{equation}
where $v$ is the speed with which the wave travels. Using the fact that $T$ corresponds to the black-hole temperature~\cite{FluidGrav,FluidGrav2,FluidGrav3}, we have
\begin{equation}
\lambda_T \sim \frac{4 \pi^2 \, c \, v}{a}
\label{aThermal}
\end{equation}
where $a$ is the surface gravity. As mentioned above, we now compare the above length scale with the excitation wavelength $\lambda_0$ of the medium surrounding the black-hole for a far-away observer. Using the red-shift relation, we have
\begin{equation}
  \frac{1}{\lambda_0}\approx \frac{1}{\lambda_e}\sqrt{\frac{\delta}{r_H}}. \label{lambda0}
 \end{equation}
where $r_H$ is the horizon radius, $\delta (\ll r_H)$ is the distance of the excitation from the horizon, and $\lambda_e$ is the excitation wavelength close to the horizon. 

Until now the discussion has been for a general black-hole. For simplicity and clarity, we now consider 4-D Schwarzschild black-holes. We get,
\begin{equation}
  \frac{\lambda_e}{\lambda_0} =  c \sqrt{\frac{\delta}{2 \, G \, M}}. \label{lambda0a}
 \end{equation}
Expressing the above equation in terms of surface gravity, we have,
\begin{equation}
 \lambda_0 = \lambda_e \frac{c}{\sqrt{2 \, \delta \, a}} \label{lambdaacccn}
\end{equation}
It is important to compare and contrast the two length-scales $\lambda_T, \lambda_0$ obtained in Eqs. (\ref{aThermal}, \ref{lambdaacccn}), respectively. First, both these length scales increase as mass of the black-hole increases. However, $\lambda_T$ increases much faster than $\lambda_0$. Second, $\lambda_0$ contains information about the shortest possible excitations. For $\lambda_T \ll \lambda_0$, for a typical macroscopic black-hole, it is impossible for an outside observer to obtain information about the Black-hole DOF. 

Let us now ask the question: "When are the two length scales comparable?". Equating \eqref{lambdaacccn} and \eqref{aThermal}, we get, 
\begin{equation}
a^c= 32\pi^4 \, \delta \, \left(\frac{v}{\lambda_e}\right)^2. \label{a}
\end{equation}
Once we know $a^c$, we can determine $T^c$ and corresponding horizon radius $r_h^c$. To determine $a_c$, we need to know $\delta$, $v$ and $\lambda_e$. Both, $\delta$ and $\lambda_e$ correspond to the smallest possible length scales which we can set to Planck length ($\ell_P$). As mentioned earlier, $\rho$ is a constant. This means that the spatio-temporal variations of $\rho$ are suppressed. In other words, the excitation of $\rho$ has a large rest-mass energy compared to its kinetic energy, i. e., 
\begin{equation}
\frac{\textrm{K.E.}}{\textrm{Rest \ mass \ energy}} = \left(\frac{v}{c} \right)^2 \sim (\frac{\ell_P}{\ell_s})^2, \label{vestm}
\end{equation}
where, $\frac{l_P}{\ell_s}$ provides the magnitude of any deviations from constant $\rho$ and usually, $\ell_s \gg \ell_P$. Substituting $\delta, v$ and $\lambda_e$ in \eqref{a}, we get, 
\begin{equation}
a^c \sim 32\pi^4 \, c^2 \frac{\ell_P}{\ell_s^2} \quad \textrm{and} \quad
r_H^c \sim \frac{1}{64 \pi^4} \frac{\ell_s^2}{\ell_P} \, .
\end{equation}
Taking $\ell_s$ to be in range $[10^{-3} m, 1 m]$, we then have $r_H^c$ in the range 
$[10^{25} m, 10^{31} m]$. The horizon radius is comparable to the Hubble horizon which is $\approx 10^{26} m$.

Let us put these results in perspective: We have related irreversible processes 
in non-equilibrium (infalling matter-energy) to thermal fluctuations (of the horizon 
DOF) in equilibrium. However, we have obtained the result without the detailed analysis 
using the fluctuation-dissipation theorem~\cite{FDT,FluidGrav3,FluidGrav4} and hence, the arbitrariness in $\ell_s$. Given the mesoscopic range of $\ell_s$, we have shown that the maximum size of the horizon radius can approximately be as large as the Hubble horizon of the Universe. This result is consistent with other arguments in the literature that the black-holes cannot grow arbitrarily large.  Based on the topological properties of FRW space-time, it was argued that black-holes could not be arbitrary large ~\cite{KerrHawk}.  For the entropy bounds to be valid, it was claimed that black-holes should be bounded from above. Specifically, Bekenstein bound suggests an upper bound to the black-hole size without which the holographic entropy bound appears to be violated~\cite{HoloBound,Veneziano}. While the analysis has been performed for Schwarzschild, the study here is general enough that this can be applied to asymptotically non-flat black-holes. Since black-holes have the highest entropy for objects of similar size, our analysis suggests that the entropy corresponding to the above horizon size is the maximum allowable entropy for any system. Our analysis poses an interesting question: Whether the entropy of the Universe \cite{Egan-Lineweaver} also has an upper bound? We plan to look into this elsewhere. 




\vspace*{10pt}

\noindent {\bf Acknowledgments} The work of SS is supported by DST-Max Planck-India Partner Group on Gravity and Cosmology, and IRCC Grant, IIT Bombay.

\end{document}